\documentclass[12pt]{iopart}

\usepackage{etoolbox}
\usepackage[utf8]{inputenc}

\makeatletter
\newcommand{\mainmatter}{%
  \setcounter{footnote}{0}%
  \patchcmd{\@makefntext}{\fnsymbol}{\arabic}{}{}%
  \patchcmd{\@thefnmark}{\fnsymbol}{\arabic}{}{}%
  \def\@makefnmark{\textsuperscript{\arabic{footnote}}}%
}
\makeatother

\usepackage{iopams}

\expandafter\let\csname equation*\endcsname\relax

\expandafter\let\csname endequation*\endcsname\relax

\usepackage{mathrsfs,mathtools,bbold}
\usepackage{amsmath,amssymb}
\usepackage{amsfonts}
\usepackage{float, cite}
\usepackage[colorlinks,linktocpage]{hyperref}
\usepackage[dvipsnames]{xcolor}
\usepackage{physics}
\usepackage{slashed}

\hypersetup{colorlinks=true,linkcolor=blue,citecolor=blue,urlcolor=blue}

\newcommand{\sqrtP}[1]{\sqrt{\smash[b]{#1}}}

\begin{document}
\title{Electric dipole moment in storage ring experiments}
\author{Peter I. Porshnev}

\address{Past affiliation: Physics Department, Belarusian State University, Minsk, Belarus}
\vspace{10pt}
\begin{indented}
\item[]July 2020
\end{indented}

\begin{abstract}
The measurement of electric dipole moment in storage rings can potentially exceed the sensitivity of tests with neutral systems. The spin dynamics under such conditions is described by the Bargmann-Michel-Telegdi equation. It can be derived in the semiclassical approximation under several assumptions one of which is the zero pseudoscalar bilinear. However, many promising extensions to the standard model consider scalar-pseudoscalar couplings which assume nonzero electron pseudoscalar. We re-derive the spin precession equation under conditions that do not assume that pseudoscalar is zero. It leads to a correction term that might be required for matching the storage ring measurements with QFT evaluations.
\end{abstract}

\vspace{2pc}
\vspace{8ex}

{\footnotesize\noindent\color{gray} This is the version of the article before peer review or editing, as submitted by an author to \emph{Physica Scripta}. IOP Publishing Ltd is not responsible for any errors or omissions in this version of the manuscript or any version derived from it. The Version of Record is available online at \url{https://doi.org/10.1088/1402-4896/abd8a3} }

\vspace{10pc}
{\footnotesize\color{gray}email: pporshnev@gmail.com}

\maketitle

\date{July 15, 2020}

\section{Introduction}
The search for electric dipole moment (EDM) of elementary particles is the focus of intense experimental and theoretical efforts, since it can potentially reveal $CP$-violating interactions outside the standard model. The unprecedently low upper bound on electron EDM was recently achieved  in experiments on neutral systems \cite{andreev_improved_2018-1}. The storage ring experiments \cite{abusaif_storage_2019-1, semertzidis_storage_2016, pretz_statistical_2020,kirch_search_2020} have the promise to reduce this limit even more, however they require the description of spin dynamics  with high accuracy \cite{khriplovich_electric_2000, rathmann_spin_2020,pretz_on_behalf_of_the_jedi_collaboration_measurement_2013}. It is the goal of this work to potentially extend the phenomenological description of spin dynamics by including an additional $CP$-violating factor that is often used in field-theoretical models, which however has been missing from typical phenomenological models.

The spin precession of relativistic electrons in electromagnetic field $F^{\mu\nu}$ is given by the Bargmann-Michel-Telegdi (BMT) equation \cite{bargmann_precession_1959} which extended version \cite{silenko_spin_2015} includes the EDM terms 
\begin{equation}\label{eedm_21}
	\dv{s^\mu}{\tau} 
	=\frac{ge}{2m} F^{\mu\nu}s_\nu+\frac{a_e e}{m}( s^\rho F_{\rho\nu}u^\nu)  u^\mu
	- 2d_e\qty( \tilde F^{\mu\nu}  s_\nu+  s^\rho \tilde F_{\rho\nu}  u^\nu u^\mu ) \,,
\end{equation}
here $s^\mu$ is the spin four-vector, $u^\mu$ is four-velocity, $a_e$ and $d_e$ are the anomalous magnetic  and electric dipole moments respectively. By measuring the spin dynamics, the goal is to resolve $d_e$ for comparison with theoretical models.

The magnitudes of $a_e$ and $d_e$ can only be derived within the field theory. The first moment is reliably evaluated within quantum electrodynamics, while $d_e$ requires the standard model or one of its extensions \cite{pospelov_electric_2005}.  Both  moments are induced  by virtual particles and thus are small with the following distinction. The small correction to electron magnetic moment is given by the second form-factor in the vertex function \cite[p. 196]{peskin_introduction_1995}. The one-loop correction is generated by emission and absorption of single virtual photon. As a result, the electron $g$-factor is modified by
\begin{equation}\label{eedm_7}
	a_e = \frac{g-2}{2} = \frac{\alpha}{2\pi}+\mathcal{O}(\alpha^2) \approx 0.0011\,,
\end{equation}
which is indeed small compared to $g=2$ however it has been confirmed in countless experiments. The electron electric moment instead is induced by virtual quarks, and requires the four-loop correction order within the standard model which makes its magnitude extremely small \cite{pospelov_electric_1991,pospelov_ckm_2014}.
Various extensions \cite{altmannshofer_electric_2020,cesarotti_interpreting_2019} that include additional $CP$-violating interactions  could lead to significantly higher values of $d_e$ which has not been observed so far.  Even if it does not seem that $a_e$ and $d_e$ are related to each other in some simple way, see expressions for $d_e$ given in \cite{khriplovich_cp_1997, bernreuther_electric_1991},  the electric moment can still be formally given by 
\begin{equation}\label{eedm_5}
	d_e  =  \lambda  \,a_e \mu_B\,,
\end{equation}
where $\mu_B$ is the Bohr magneton, and the dimensionless coefficient $\lambda$ combines model-specific mass ratios, mixing angles, and couplings. It was proposed based on purely dimensional grounds \cite{engel_electric_2013}.

The Bargmann-Michel-Telegdi  equation \eqref{eedm_21} can be derived or justified in three different ways: by using the heuristic arguments \cite{berestetskii_quantum_1982, fukuyama_derivation_2013}, with the Foldy–Wouthuysen transformation \cite{silenko_quantum-mechanical_2005} or within the semiclassical approximation to the Dirac equation \cite{rafanelli_classical_1964}. All derivations agree with each other and lead to the well known form \eqref{eedm_21}. It has been tested in multiple experiments, and there is no doubt  in its validity under conditions where the equation is applicable.  This work is yet another derivation of the BMT equation with the EDM term which leads to the form \eqref{eedm_21} enhanced with an otherwise small correction.  Such a small correction that has safely been ignored so far can now become important, since the modern experiments strive to achieve a higher degree of accuracy.

While the heuristic derivation is based on several a priori assumptions, the semiclassical derivation strictly outlines them. One of these assumptions is the zero pseudoscalar bilinear \cite{rafanelli_classical_1964}; it is the only derivation known to us that makes this assumption explicitly. The other derivations, which are cited above, do not mention the pseudoscalar coefficient, hence they implicitly assume that it is zero. It was a reasonable assumption when the equation was derived in \cite{rafanelli_classical_1964}; the BMT equation with the EDM terms works reliably in many cases \cite{khriplovich_feasibility_1998} where the pseudoscalar is assumed to be negligible. However, several theoretical groups now consider interaction terms  in field Lagrangians  that include nonzero pseudoscalar $q=i\bar \psi\gamma^5\psi\ne 0$.  In other words, this  disconnect can be stated as follows. From the one hand, the coefficient $d_e$ is derived from $CP$-violating models \cite{engel_electric_2013,pospelov_ckm_2014} which consider scalar-pseudoscalar terms or forms that are reduced to them by Fierz identities. It means both $d_e$ and $q$ might be nonzero in such models. From the other hand, such an electric dipole moment $d_e$ is used in the BMT equation \eqref{eedm_21} which implicitly assumes that $q=0$. The resolution of this disconnect is the main result of this work. 

It is important to stress again that the traditional BMT equation is essentially correct. Our derivation only adds a correction to the EDM coefficient $d_e$ itself, while keeping the same functional form \eqref{eedm_21} that was derived in \cite{berestetskii_quantum_1982, fukuyama_derivation_2013, silenko_quantum-mechanical_2005, rafanelli_classical_1964}, and many other sources.  Since the functional form of BMT equation  \eqref{eedm_21} does not change, we can immediately apply its other forms, including the Thomas-BMT one \cite{silenko_spin_2015,fukuyama_systematic_2018}, for comparison with experiments in laboratory frames. However, our results show that the coefficient $d_e$ must be adjusted for the otherwise small contribution under certain conditions. Since the EDM itself seems to be very small, based on the latest data \cite{andreev_improved_2018-1}, it is important to describe the spin precession dynamics as accurately as possible. 

We re-derive the spin precession equation under conditions that do not assume zero pseudoscalar. Within the semiclassical approximation to the Dirac equation, the observable electric moment $d_e^{\text{exp}}$ is found to be
\begin{equation}\label{eedm_4}
		d_e^{\text{exp}} \sim 2d_e - q\, a_e \mu_B\,.
\end{equation}
It means that the BMT equation does not change its form \eqref{eedm_21}, however the EDM coefficient $d_e$ is shifted by the amount that is proportional to $q$.  The quantity $d_e^{\text{exp}}$ is directly related to the frequently used parameter $\eta$ \cite{fukuyama_derivation_2013, kirch_search_2020, semertzidis_storage_2016} which is  $\eta = \frac{2 d_e m }{e s} = \frac{2 d_e}{\mu_B}$ if $s=1/2$. Remember $d_e$ is just the coefficient from field Lagrangians.  If $q=0$, the observable $d_e^{\text{exp}} = \eta \mu_B$. The second term in \eqref{eedm_4} has the correct dimension which is similar to \eqref{eedm_5}; it also properly transforms in discrete symmetry transformations. The adjusted EDM coefficient $d_e^{\text{exp}}$ must be used in the BMT equation \eqref{eedm_21} instead of $2d_e$ for $q \ne 0$.  If both terms \eqref{eedm_4} have comparable contributions into the observable dipole moment $d_e^{\text{exp}}$, the interpretation of storage ring experiments has to take it into account.  

Remarkably, the similar situation was in fact already discussed for atomic and molecular EDM experiments \cite{flambaum_sensitivity_2020, pospelov_ckm_2014}.\footnote[7]{\footnotesize \color{gray}Credit to Dr.~Stadnik for pointing  this analogy to us upon reviewing the manuscript.} Provided the electrons are coupled to nucleons via the pseudoscalar-scalar term, its contribution cannot be distinguished from the $d_e$ proper in the case of single species experiments \cite{pospelov_ckm_2014}. Similarly, for the case of scalar-pseudoscalar semileptonic operators \cite{flambaum_sensitivity_2020} used in describing experiments with paramagnetic atoms \cite{andreev_improved_2018-1}, these two contributions cannot be distinguished from each other if the energy shift is measured in a single experiment. Ditto for the case of electron-axion coupling \cite{stadnik_improved_2018}. However, we do not think that the mixing of contributions stemming from $d_e$ proper  and the pseudoscalar $q$ has ever been discussed for storage ring experiments. More, the semiclassical approximation allowed us to obtain the observed $d_e^{\text{exp}}$ in the closed form \eqref{eedm_4} which we believe is new.

The paper is organized as follows. In the next section, we derive the spin equation for Dirac electron within the WKB approximation, and obtain the expression for $d_e^{\text{exp}}$. The results  will be discussed in the third section. 

\section{Semiclassical approximation}
We do not invent any new methods here, and strictly follow the derivation \cite{rafanelli_classical_1964} which is based on the WKB expansion of Dirac wave function. It is straightforwardly extended to include the EDM term. The critical point comes at the very end where \cite{rafanelli_classical_1964} explicitly sets the pseudoscalar to zero. We do not make this assumption in the agreement with the proposed mechanisms of CP violation which we cited above.

The additional electron moments are represented by two terms in the Lagrangian 
\begin{align}\label{eedm_3}
	&\frac{a_ee}{4m}F_{\mu\nu}\bar\psi\sigma^{\mu\nu}\psi\,,	&i \frac{d_e }{2}F_{\mu\nu}\bar\psi\sigma^{\mu\nu}\gamma^5\psi\,.
\end{align}
 These terms can be postulated and added to the Dirac Lagrangian or obtained in effective field theories \cite{pospelov_electric_2005}. Their universal form covers any spin one-half fermion including nucleons \cite{safronova_search_2018,flambaum_sensitivity_2020}, and can also be seen in extensions to the standard model \cite{altmannshofer_electric_2020}.

The semiclassical derivation \cite{rafanelli_classical_1964} is extended to include the EDM term and nonzero $q$. Correspondingly to \eqref{eedm_3}, two terms are added to the Dirac equation 
\begin{equation}\label{chFields:eq1967}
	\qty[i \slashed \partial -  e \slashed A_\mu-m - (\frac{a_e e}{2m}+id_e \gamma^5)\frac{\sigma^{\mu\nu}}{2}F_{\mu\nu}] \psi = 0\,, 
\end{equation}
where the signs of terms with moments are selected to match the extended BMT equation from \cite{fukuyama_derivation_2013}.
The Dirac equation reduces to the classical limit in the eikonal approximation 
\begin{equation}\label{chFields:eq1966}
	\psi(x) = \phi e^{i S(x)}\,,
\end{equation}
which assumes the constant $\phi$. To obtain the BMT equation, we need to refine
\eqref{chFields:eq1966} by allowing $\phi(x)$ to be a slow-changing function of coordinates. Substituting \eqref{chFields:eq1966} into \eqref{chFields:eq1967}, we obtain in the leading order
\begin{equation}
	\Big[\gamma^\mu(\partial_\mu S + eA_\mu)+m\Big]\phi = 0 \,,
\end{equation}
where the similar equation is obtained for the adjoint bispinor. They both lead to the Hamilton-Jacobi equation where the phase derivatives $\partial_\mu S$ are associated with the classical Hamiltonian and conjugate momentum
\begin{equation}\label{eedm_6}
	p_\mu = - (\partial_\mu S + eA_\mu) = m u_\mu\,.
\end{equation}
Since the Hamilton-Jacobi equation  is  one of many equivalent forms that describe the classical motion, the traditional four-dimensional equation of motion follows 
\begin{equation}\label{chFields:eq1979}
			m\dv{u_\mu}{\tau}=  e  F_{\mu\nu}u^\nu\,,
\end{equation}
where $u_\mu$ the unit-norm four-velocity. The assignment \eqref{eedm_6} will be further validated below by deriving \eqref{chFields:eq1979} from the next approximation order to the Dirac equation. 

Next, squaring \eqref{chFields:eq1967} with the adjoint operator
\begin{equation}
	i \slashed \partial -  e \slashed A_\mu+m + (\frac{a_e e}{2m}+id_e \gamma^5)\frac{\sigma^{\mu\nu}}{2}F_{\mu\nu}\,, 
\end{equation}
we obtain the second order equation 
\begin{equation}
	\Big[(i\partial-eA)^2 -m^2- ma_{\mu\nu} \sigma^{\mu\nu} -2ib_{\mu\nu} \gamma^\nu(i\partial^\mu-eA^\mu) \Big]\psi=0\,,
\end{equation}
where two tensor coefficients are defined as
\begin{align}
	&a_{\mu\nu}=\frac{g e}{4m}F_{\mu\nu}-d_e\tilde F_{\mu\nu}\,,	&b_{\mu\nu}=\frac{a_e e}{2m}F_{\mu\nu}-d_e\tilde F_{\mu\nu}\,.
\end{align}
This equation is derived from  the original Dirac equation in the limit of small and slow-changing fields. Since the kinetic term is scalar now, the spin terms are explicit. Next steps are to use \eqref{chFields:eq1966} and expand the kinetic term. Ignoring the second derivatives of $\phi(x)$, we obtain
\begin{equation}\label{chFields:eq1972}
	2 p^\mu \partial_\mu\phi +\phi \partial_\mu p^\mu+ ima_{\mu\nu}\sigma^{\mu\nu} \phi -2b_{\mu\nu} p^\mu\gamma^\nu  \phi =0\,.
\end{equation}
It is the main equation in the first approximation order. Several quasi-classical quantities can be derived from it by forming corresponding bilinears.

In addition to $q$, four more real-valued quantities \cite[p. 102]{berestetskii_quantum_1982} with distinct transformation properties can be constructed from $\psi$
\begin{gather}
\begin{aligned}\label{relQM:eq59}
	&r &&= \bar{\psi}\psi\,,	&&\qquad B^{\mu\nu}&&=\bar{\psi} \sigma^{\mu\nu}\psi\,,\\[1ex]
	&k^\mu &&= \bar{\psi}\gamma^\mu\psi\,,	&&\qquad w^\mu &&= \bar{\psi}\gamma^\mu\gamma^5\psi\,.
\end{aligned}
\end{gather}
The products of bilinears are interrelated by the Fierz identities
\begin{gather}
\begin{aligned}\label{chFields:eq976}
	&k^\mu k_\mu = -w^\mu w_\mu &&= r^2 + q^2=n^2\, , &\qquad k^\mu w_\mu = 0\, , \\[1ex]
	&k^\mu w^\nu - w^\mu k^\nu &&= q B^{\mu\nu}+ r \tilde B^{\mu\nu}\, .
\end{aligned} 
\end{gather} 
Many more similar identities can be found. For example, we can reverse the last identity in \eqref{chFields:eq976} to obtain
\begin{equation}\label{chFields:eq1982}
	(r^2+q^2)B_{\mu\nu} = q(k_\mu w_\nu - w_\mu k_\nu) - r \varepsilon_{\mu\nu\rho\sigma} k^\rho w^\sigma\,.
\end{equation}
They are especially simple if the pseudoscalar $q$ is zero.

To link the vector bilinears with corresponding classical quantities, we must normalize them as
\begin{align}
	&p_\mu=\frac{m}{n}\, \bar{\phi}\gamma_\mu\phi\,, 
	&s_\mu= \frac{1}{n}\,  \bar{\phi}\gamma_\mu\gamma^5\phi\,,
\end{align}
where $n=\sqrtP{r^2+q^2}$.  The current conservation becomes
\begin{equation}\label{chFields:eq1975}
	\partial_\mu(\bar{\psi}\gamma^\mu\psi)\sim\partial_\mu(p^\mu n )\sim\partial_\mu(u^\mu n)=0\,.
\end{equation}
These assignments will be checked for consistency with the other equations. 

The next step is to add together \eqref{chFields:eq1972} and its adjoint after we multiply them with $\bar\psi \Gamma$ and $\Gamma \psi$ respectively; here $\Gamma=\{1, \gamma^\mu, \gamma^\mu\gamma^5\}$. The first $\Gamma$-factor leads to the continuity equation
\begin{equation}
		p^\mu \partial_\mu r +\partial_\mu(r  p^\mu)-4b_{\mu\nu} p^\mu\bar{\phi}\gamma^\nu  \phi
		\approx \qty(1+\frac{q^2}{r^2})p^\mu \partial_\mu r -\frac{q}{r}p^\mu \partial_\mu q\approx p^\mu \partial_\mu r= 0
\end{equation}
where we ignored the derivatives of $q$ and assumed that $q^2\ll r^2$. The density $r=\bar{\phi}\phi$ does not change along the particle trajectory 
\begin{equation}
	\frac{1}{m}p^\mu \partial_\mu r = u^\mu \partial_\mu r= \dv{r}{\tau}=0\,,
\end{equation}
which is expected for the classical particle. It also serves as the definition of total derivative over proper time in the classical limit of Dirac physics.

The second $\Gamma$-factor leads to
\begin{equation}
	2 m \qty[ u^\mu\partial_\mu(u^\rho n)-2 n a^{\rho\nu} u_\nu -2r b^{\mu\rho}u_\mu]=0\,.
\end{equation}
Canceling the normalization factor $r\approx n$ that is constant along the particle trajectory, the above equation becomes
\begin{equation}
	\dv{u^\rho}{\tau} = 2( a^{\rho\mu} u_\mu- b^{\rho\mu}u_\mu)
		= \frac{e(g-2a_e)}{2m}F^{\rho\mu} u_\mu= \frac{e}{m}F^{\rho\mu} u_\mu\,.   
\end{equation} 
It matches the classical equation \eqref{chFields:eq1979} if $a_e$ is defined as in \eqref{eedm_7}. Remarkably, both anomalous magnetic and electric dipole moments do not enter the equation of motion in the first approximation order. They will appear in higher order equations \cite{pomeransky_equations_1999} which are outside the scope of this work. However, both moments impact the spin precession dynamics in the first approximation order; the similar comment was made previously in \cite{silenko_quantum-mechanical_2005}.    

Lastly, the third $\Gamma$-factor leads to the spin equation
\begin{equation}
	u^\mu\partial_\mu(s^\rho n)-2 na^{\rho\nu} s_\nu+\frac{2}{r} b_{\mu\nu} u^\mu (n^2u^\rho s^\nu -  q\bar{\phi}\sigma^{\rho\nu}\phi)   =0\,,
\end{equation}
which after canceling the normalization factor becomes
\begin{equation}\label{chFields:eq1981}
		\dv{s^\rho}{\tau}  =\frac{ge}{2m} F^{\rho\nu}s_\nu + \frac{a_e e}{m}( s^\mu F_{\mu\nu}u^\nu)  u^\rho 
		- 2 d_e\qty(\tilde F^{\rho\nu}  s_\nu+ s^\mu \tilde F_{\mu\nu}  u^\nu u^\rho ) + \frac{2q}{r^2} b_{\mu\nu} u^\mu  (\bar{\phi}\sigma^{\rho\nu}\phi)\,.
\end{equation}
It matches the BMT equation \cite{bargmann_precession_1959} if $d_e$ and $q$ are set to zero. If only $q$ is set to zero, \eqref{chFields:eq1981} matches its extended version \cite{fukuyama_derivation_2013} which includes the EDM terms.

The last term in \eqref{chFields:eq1981} is new. Up to this point, we have followed the derivation \cite{rafanelli_classical_1964} with only one difference: the equations have been updated with the EDM terms. Here comes the critical juncture. Instead of setting $q$ to zero as it was done in \cite{rafanelli_classical_1964}, we use the Fierz identity \eqref{chFields:eq1982} to transform the last term in \eqref{chFields:eq1981}. It leads to the following equation 
\begin{multline}\label{chFields:eq1984}
	\dv{s^\rho}{\tau} 	=\qty(\frac{ge}{2m}+2d_e\frac{q}{r}  ) F^{\rho\nu}s_\nu
	+\qty[\frac{a_ee}{m}(1-\frac{q^2}{r^2})+2d_e\frac{q}{r} ] s^\mu F_{\mu\nu}u^\nu u^\rho\\[1ex]
	- \qty( 2 d_e- \frac{a_ee}{m} \frac{q}{r}) \tilde F^{\rho\nu}  s_\nu
	- \qty[2 d_e(1 -\frac{q^2}{r^2}) -\frac{a_e e }{m}\frac{q}{r} ] s^\mu \tilde F_{\mu\nu}  u^\nu u^\rho  \,.
\end{multline}
where we see that $q$ adds contributions to all terms in the semiclassical spin equation. Since the magnetic moment  is known with high precision, upper bounds can be placed on new terms added to the magnetic moment terms in \eqref{chFields:eq1984}. Assuming the calculations \cite{aoyama_revised_2018} include neither of these additional terms, the upper bounds are 
\begin{equation}\label{chFields:eq1989}
	d_e \frac{q}{r} < \delta \qty(\frac{a_e e}{2m})= 2\cross10^{-22} e\,cm\,,\qquad	\qty(\frac{q}{r})^2 < 10^{-11}\,,
\end{equation} 
which means that $\abs{q/r}<4\cross 10^{-6}$; we expect $r\approx 1$. This upper bound comes from the uncertainty in measuring the electron AMM. Since the experimental bound \cite{andreev_improved_2018-1} on $d_e$ is much tighter than \eqref{chFields:eq1989}, the pseudoscalar correction can be ignored for the magnetic terms. However, this conclusion is only valid for electrons. The derivation is equally applicable to any spin one-half massive particle; hence, heavier fermions might acquire meaningful corrections to the magnetic moment terms. 

The electron spin equation then becomes
\begin{equation}\label{chFields:eq1990}
	\dv{s^\mu}{\tau} 
		=\frac{ge}{2m} F^{\mu\nu}s_\nu+\frac{a_e e}{m}( s^\rho F_{\rho\nu}u^\nu)  u^\mu
		- d_e^{\text{exp}}\qty( \tilde F^{\mu\nu}  s_\nu+  s^\rho \tilde F_{\rho\nu}  u^\nu u^\mu )  \,,
\end{equation}
where the adjusted EDM is 
\begin{equation}\label{eedm_10}
	d_e^{\text{exp}} = 2 d_e -\frac{q}{r} \frac{a_e e }{m} \,.
\end{equation}
The modified BMT equation \eqref{chFields:eq1990} collapses to its traditional form \cite{fukuyama_derivation_2013} if $q=0$. As a quick sanity check, the additional term in \eqref{eedm_10} has the correct dimension, and it also properly transforms in discrete symmetry transformations, including the parity one.

\section{Discussion}
Comparing \eqref{chFields:eq1990} with \eqref{eedm_21}, we see that our derivation yields the same functional form \eqref{eedm_21} of the BMT equation that was derived in \cite{berestetskii_quantum_1982, fukuyama_derivation_2013, silenko_quantum-mechanical_2005, rafanelli_classical_1964}, and many other sources. 
The immediate benefit is we can immediately apply its other forms, including the Thomas-BMT one \cite{silenko_spin_2015,fukuyama_systematic_2018}, for comparison with experiments in laboratory frames. However, the leading coefficients in \eqref{eedm_21} do acquires the constant correction terms. Since these additional terms are frame-independent, the functional dependencies of spin precession on velocities and fields in laboratory frames remain intact. These corrections are too small for the  electron magnetic moments, and thus are dropped from the first two terms in \eqref{chFields:eq1990}.
However, the correction to $d_e$ can be significant. 

Specifically, we showed that the observable related to EDM includes the additional contribution connected to $a_e$ if the pseudoscalar is nonzero.  This conclusion is valid for any frame including the rest one
\begin{equation}
	\dv{\va{s}}{t} = \frac{ge}{2m} \va{s}\cross\va{B} + d_e^{\text{exp}}\,\va{s}\cross\va{E}\,,
\end{equation}
since the pseudoscalar correction enters two terms in \eqref{chFields:eq1990}.
The original intent has been to derive the extended BMT equation that applies to  the storage ring experiments \cite{rathmann_electric_2019,abusaif_storage_2019-1,semertzidis_storage_2016} with relativistic particles. The obtained expression \eqref{eedm_10} ties together the  Lorentz scalars only; it does not include velocities or other frame-dependent quantities. It should be valid then in any frame, including the rest one. We have already commented that the mixing of $d_e$ and terms connected to scalar-pseudoscalar coupling was also found for atomic/molecular systems \cite{pospelov_ckm_2014, flambaum_sensitivity_2020}. It remains to be seen whether the closed form \eqref{eedm_10} adds anything new to the case of non-relativistic and neutral systems. We believe that it might be the case, since the derivation is based on the Dirac equation which is valid in both relativistic and non-relativistic cases. Since non-relativistic and neutral systems are outside the scope of this work, it will be investigated in detail elsewhere.

Even if $d_e$ proper and $q$ are both Lorentz-invariant quantities, they are very different from each other by definition. The intrinsic dipole moment is considered the fundamental property of elementary particle as its charge, spin, magnetic moment, and mass. In general, the electron pseudoscalar $i\bar\psi\gamma^5\psi$ is a dynamic quantity that depends on anything that influences its wave function in a certain way. Since the observable $d_e^{\text{exp}}$ inherits the dynamic dependence from $q$, it could potentially vary in experiments, depending on $q$. It  means that the form \eqref{eedm_10} could also help in connecting experiments with theoretical models that study the time-varying EDM \cite{graham_axion_2011,stadnik_axion-induced_2014, budker_proposal_2014, abel_search_2017-1}. 

What is really measured in EDM experiments,  $d_e$ proper or $d_e^{\text{exp}}$?  The coefficient $d_e$ is just the constant from Lagrangian, while $d_e^{\text{exp}}$  is the outcome of evaluating the matrix element. In transitioning from \eqref{chFields:eq1981} to \eqref{chFields:eq1984}, we saw that $q$ intermixes terms in the semiclassical spin equation. Anyway, since everything we measure must be given by matrix elements, the answer is  clear.   Could two terms in \eqref{eedm_10} cancel each other, thus leading to tiny $d_e^{\text{exp}}$? It is known \cite{bernreuther_electric_1991, cesarotti_interpreting_2019} that the dipole moment interaction flips the electron chirality which is related to pseudoscalar quantities. If nonzero $d_e$ and $q$ are really two faces of the same symmetry-violating mechanism, both terms might have comparable contributions into $d_e^{\text{exp}}$. Depending on their relative signs, it might have the dramatic impact on the EDM search. Even if a given extension to the standard model predicts a nonzero $d_e$ proper, the observable $d_e^{\text{exp}}\approx 0$ will reveal nothing  if two contributions cancel each other. Still, such a case does not defeat the main quest of searching for symmetry violations, since the pseudoscalar is nonzero. The experiments would have to be adjusted though. The least exciting outcome is if both $d_e$ and $q=i\bar\psi\gamma^5\psi$ are negligible individually.  

Assuming instead that $q$-term is the dominant contribution into $d_e^{\text{exp}}$, we can find the upper bound on the electron pseudoscalar by using the data \cite{andreev_improved_2018-1}  
\begin{equation}\label{eedm_9}
	\frac{q}{r} \frac{a_e e }{m} =\frac{q}{r}(10^{-3})\,( 2\cross 10^{-11})e\,cm < 10^{-29} e\,cm\,.
\end{equation} 
It means that $\abs{i\bar\psi\gamma^5\psi}<5\cross 10^{-16}$. It is the extreme case; most probably, this additional term adds a correction to $d_e$ proper to yield the observable value $d_e^{\text{exp}}$. Hopefully, the obtained results are found  useful in the interpretation of EDM experiments including the ones that have recently  been announced \cite{abe_new_2019-1, ho_new_2020}.

Concluding, the known derivations of the BMT equation \cite{bargmann_precession_1959, berestetskii_quantum_1982, fukuyama_derivation_2013, silenko_quantum-mechanical_2005}, including this work, all agree with each other if $q=0$. For all practical purposes, if a CP violating model predicts a significantly nonzero value of $d_e$ and very small value of $q$, no correction is required to the known spin precession equation \eqref{eedm_21}. 

\ack

We thank Y.~Stadnik and D.~Budker for reviewing the manuscript and providing valuable suggestions.

\section*{References}

\bibliographystyle{unsrt}

\end{document}